\documentclass[twocolumn,showpacs,prc]{revtex4}
\usepackage[dvips]{graphicx}
\usepackage{amsmath,amssymb}
\usepackage{dcolumn}
\begin{document}
\title{
Relativistic hydrodynamical model in the presence of long-range correlations
}
\author{T. Osada}
\email{osada@ph.ns.tcu.ac.jp}
\affiliation{
Department of Physics, School of Liberal Arts, Tokyo City University,
Tamazutsumi 1-28-1, Setagaya-ku, Tokyo 158-8557, JAPAN}
\date{\today}
\begin{abstract}
Effects of dynamical long-range correlations over a fluid cell
size scale on a relativistic fluid are discussed. It is shown that
such correlations among the fluid elements introduced into
hydrodynamical model induce some weak dissipation and viscosity
into the fluid. The influence of the long-range correlations on
the entropy current is also discussed.
\end{abstract}
\maketitle

\section{Introduction}\label{sec:Introduction}
Recently, the so called ridge phenomenon has been observed at RHIC
\cite{PutschkeJPhysG34} (particle yield enhancement in narrow
azimuthal angle window $\Delta \phi$ but broad pseudo-rapidity
window $\Delta \eta$). It is a jet-related long range correlation
in rapidity. For us it is particularly interesting that the
so-called {\it soft ridge} phenomenon, a similar longitudinal
rapidity correlation but without jet trigger, is found in
collisions of higher centrality events
\cite{AdamsJPhysG32,DaugherityJPhysG35}. Although there is no
theoretical model that can quantitatively reproduce all
experimental data related to this phenomena, several models have
been proposed aiming at the explanation of ridge data
\cite{GavinPRC79,DumitruNPA810,Lappi0909.0428,Peitzmann0903.5281,Takahashi0902.4870,Gavin0806.4366,Hama0911.0811}.

Actually, correlations over several rapidity units can only occur
at the earliest stages of heavy-ion collision, when different
rapidity regions are still causally connected
\cite{DumitruNPA810,Peitzmann0903.5281}. It means that the origin
of this phenomenon must be placed almost instantaneously after the
collision of two nuclei \cite{GavinPRC79,DumitruNPA810}, namely
before or around the {\it initial stage} of the hydrodynamical
evolution of the fireball created by the relativistic heavy-ion
collisions. On the other hand, observation of correlations is
related to period after freeze out, what means that the long-range
correlations can survive the whole stage of hydrodynamical
evolution of the matter. Since little attention has been given to
this problem so far it is therefore naturally to ask what are the
influence of the long-range correlations on the fluid dynamical
evolution because, this will be the subject we shall investigate
here. Because such dynamical long-range correlations may cause
weak dissipation and viscosity in the fluid we propose a kind of
extreme model for the relativistic dissipative hydrodynamics with
small dissipative and viscous terms, which originate entirely from
the long-range correlations among the fluid elements.
\\

A few words on hydrodynamical models used in particle production
processes are in order here. The ideal fluid dynamics
\cite{HamaBrazJPhys35,HuovinenPLB503,KolbPRC67,HiranoPRC65,HeinzJPhysG30}
have successfully reproduced experimental results obtained at
Relativistic Heavy-Ion collider (RHIC) (for a review, see for
example, \cite{Hirano0808.2684}). This success brought us a new
understanding of the matter created at the RHIC. It turned out
that this matter behaves more like an ideal fluid rather than the
weakly interacting parton plasma \cite{Munzinger0801.4256}. It is
thought that this is because the mean free path of the particles
composing it is small comparing to the system size or to the
typical scale in the fluid dynamics (strongly interacting quark
gluon plasma, sQGP
\cite{TDLeeNPA750,GyulassyNPA750,ShuryakNPA750}). However, there
are noticeably differences between predictions from the ideal
fluid model and the corresponding data for the flow parameters
obtained in
\cite{STARCollabPRC72,PHENIXCollabPRL91,PHOBSCollabPRC72}
\footnote{Ideal hydrodynamical models (without fluctuations in the
initial condition)
\cite{HuovinenPLB503,KolbPRC67,HiranoPRC65,HeinzJPhysG30} seem to
work for minimum bias data on elliptic flow parameter $v_2$
\cite{OllitraultPRD46}  but not for data on its centrality
dependence observed in Au+Au collisions at $\sqrt{s_{\rm NN}}=200$
GeV \cite{STARCollabPRC72}. They can also reproduce the mass
ordering and magnitude of $v_2$ for the different particles in the
region up to 2 GeV/$c$, but they fail to reproduce then for
$p_{\rm T} > 2$ GeV/$c$
\cite{PHENIXCollabPRL91,PHOBSCollabPRC72}.}. Hence, some revisions
to the ideal fluid picture turned out to be necessary and
dissipation and viscosity effects have been introduced to the
perfect fluid. This caused some problems with the proper
formulation of the relativistic dissipative fluid model
\cite{MurongaPRC69,ChaudhuriPRC74,HeinzPRC73,BaierPRC73,
DumitruPRC76,KoidePRC75,TsumuraPLB646,SongPRC77,HuovinenPRC79}. In
a simple relativistic extension of the Naiver-Stokes equation
(so-called first order theories \cite{Eckart1940, DanielewiczPRD31}), it
is known that the causality is violated
\cite{IsraelAnnPhys100,IsraelAnnPhys118} and the solution of the
equation is not stable against small perturbations
\cite{HiscockAnnPhys151,HiscockPRD35}. Over the past few years a
considerable number of studies have been made aiming at the
solution of these problems (see, for example
Ref.\cite{Romatschke0902.3663,DenicolJPhysG35,Pu0909.906}). \\

The purpose of this paper is to formulate the relativistic
imperfect hydrodynamics including  the effect of long-range
correlations and to find its connections with the usual
dissipative hydrodynamics. Note that, in the presence of the
long-range correlations over the thermal equilibrium scale, the
microscopic state of the particle that composes the fluid does not
obey the usual Boltzmann-Gibbs statistics any longer
\cite{Sherman-Rafelski2004}. Also thermodynamic functions such as
the energy density and pressure among the separate fluid cells
will be affected. The state in the presence of the long-range
correlations, which is generally different from the usual
equilibrium state (it is a kind of a non-equilibrium state), can
be the so-called stationary state near the equilibrium. It
approaches to the equilibrium state with a finite (or infinitely
long) relaxation time under a non-equilibrium thermodynamical
constraint which may make a kind of narrow valley of the free
energy generated by dynamical correlations \cite{Kodama0812.4748}.
This relaxation process depends on the initial state selected in
the valley of the free energy. Such dependence on the initial
state for the relaxation processes can be then regarded as a kind
of memory effect induced by the long-range correlations.

The relaxation process of the stationary state is called {\it
pre-thermalization} \cite{Kodama0812.4748} (see also
\cite{GavinNPB351}) and leads to Tsallis's statistics
\cite{Tsallis1988}. It is usually considered to be important for
describing correlated system and appears when the phase space is
reduced by the correlations among constituents of the system. In
\cite{OsadaPRC77} it is shown that there is a certain
correspondence between the fluid dynamics based on the Tsallis
statistics and the dissipative fluid dynamics based on the usual
Boltzmann-Gibbs statistics. Thus, the dynamical long-range
correlations may result in some dissipative contribution in the
relativistic fluid.\\

It should be stressed at this point that situation considered
here is rather special. Our model assumes that, because of action
of the long-range correlations, some stationary, near equilibrium,
state of fluid is formed replacing the usual equilibrium state.
The natural question is how real is such assumption. So far there
is no hard evidence for creation of such state. On the other 
hand, there is growing evidence of a power law-like behavior of
the transverse momentum spectra \cite{OsadaPRC77,STAR_PRL97} in
the mid $p_{\rm T}$ region (2$\sim$6 GeV/c) and this may indicate
the existence of such stationary state \cite{OsadaPRC77}. The
necessary condition for the existence of such pre-thermalization
state in the matter created in the relativistic heavy-ion
collisions (for example at RHIC) is that its relaxation (or
equilibration) time is long enough comparing with the typical
hydrodynamical time scale, such as $L_{\rm hydro}/c_s$, where
$L_{\rm hydro}$ is some typical length scale of fluid and $c_s$ is
sound velocity. All these calls for some very detail
considerations which are, however, outside of the limited scope of
the present paper. Here we consider therefore only a kind of {\it
thought experiment} in which appearance of the stationary state
due to the long range correlations is assumed and its consequences
investigated in detail.\\

The long-range correlation influence not only
the microscopic state but also the macroscopic state of the fluid
such as velocity vectors. Therefore, instead of specifying the
type of the long-range correlations, we shall assume that all
information on them are included in a flow velocity vector field,
$u^{\mu}(x)$, and in the non-equilibrium thermodynamic functions
like local energy density, $\varepsilon(x)$, and pressure, $P(x)$
(here $x$ denotes a space-time four vector).\\

To quantitatively investigate the effects of the long-range
correlations (LRC) on the relativistic fluid, we shall propose
{\it a thought experiment} assuming that it is possible to switch
on and off the LRC for a perfect fluid in such a way as not to
break the causality. If one can turn on the switch of the LRC that
acts on the perfect fluid at a proper time $x^0=\tau_{\rm on}$,
then, after $\tau_{\rm on}$, the fluid starts to expand as a
dissipative fluid. We assume that solution of the perfect fluid,
which is obtained in the case without LRC, is known for all
space-time points (i.e., by solving ideal fluid dynamics after
$\tau=\tau_{\rm on}$ we know the energy density, $\varepsilon_{\rm
eq}(x)$, conserved charge density, $n_{\rm eq}(x)$, flow velocity,
$U^{\mu}(x)$, and equation of state $P_{\rm eq}(x)=P_{\rm
eq}(\varepsilon_{\rm eq},n_{\rm eq})$ for all space-time point $x$
\cite{PeschanskiPRC80, CsorgoPLB565}. It is then assumed  that to
know evolution of changes (defined as $\Lambda(x)\equiv
\varepsilon-\varepsilon_{\rm eq}$, $\Pi(x)\equiv P-P_{\rm eq}$,
$\delta n(x)\equiv n-n_{\rm eq}$ and $\delta u^{\mu}\equiv
u^{\mu}(x)-U^{\mu}(x)$) is equivalent to solving equations of
resulting dissipative hydrodynamics. As will be seen below, since
$\Lambda$, $\Pi$ and $\delta n$ can be expressed as a function of
the enthalpy, $h_{\rm eq}\equiv \varepsilon_{\rm eq}+P_{\rm eq}$,
$n_{\rm eq}$, $U^{\mu}$ and $\delta u^{\mu}$, in order to
investigate the dissipative effects of the LRC is sufficient to
know only the evolution of $\delta u^{\mu}(x)$. Hereafter, we
shall call the set of solutions of the ideal hydrodynamical
equations, $\varepsilon_{\rm eq}(x)$, $P_{\rm eq}(x)$, $n_{\rm
eq}(x)$ and $U^{\mu}(x)$, the {\it reference fluid fields} (RFF)
for the corresponding dissipative fluid field. One may also define
RFF by switching-off the LRC for a given fluid with dissipation
that originated from LRC. \\

The paper is organized as follows. In Sec. II, we briefly review
the relativistic fluid dynamics. Then, in Sec. III, we introduce
the RFF which satisfy ideal hydrodynamics and we express the
energy-momentum tensor in terms of $\delta u^{\mu}$ and RFF. In
Sec.IV equation on $\delta u^{\mu}$ is found which agrees with the
relativistic dissipative hydrodynamic equations. Sec. V contains
the off-equilibrium entropy current corresponding to our model. We
close with Sec. VI containing the summary and some further
discussion.

\section{The relativistic fluid dynamics}\label{sec:02}
Basic equations of the relativistic hydrodynamics for arbitrary
dissipative fluid consist of by local conservation laws for the
energy-momentum and conserved charges. In addition to these local
conservation laws, the second law of thermodynamics is required.
These equations are generally expressed by the covariant
derivatives  \footnote{ The covariant derivative is defined by the
Christoffel symbol $\Gamma^{\nu}_{\lambda\mu}\equiv
\frac{1}{2}g^{\nu\sigma}(\partial_{\mu}g_{\sigma\lambda}+
\partial_{\lambda} g_{\sigma\mu}-\partial_{\sigma} g_{\lambda\mu})$.
Hence, we have
$T^{\mu\nu}_{;\mu}=\partial_{\mu} T^{\mu\nu} +\Gamma^{\mu}_{\sigma\mu} T^{\sigma\nu}
+\Gamma^{\nu}_{\sigma\mu}T^{\mu\sigma}$ and
$N^{\nu}_{;\mu}\equiv \partial_{\mu}N^{\nu}+\Gamma^{\nu}_{\lambda\mu}N^{\lambda}$.}
of the energy-momentum tensor, $T^{\mu\nu}(x)$, the conserved
charge current, $N^{\mu}(x)$, and the entropy current
$S^{\mu}(x)$;
\begin{subequations}\label{eq:01}
\begin{eqnarray}
 T^{\mu\nu}_{;\mu} &=&0, \label{eq:01a} \\
 N^{\mu}_{;\mu} &=&0,    \label{eq:01b} \\
 S^{\mu}_{;\mu} &\ge& 0. \label{eq:01c}
\end{eqnarray}
\end{subequations}
The general form of the energy-momentum tensor used here is given
by
\begin{eqnarray}
T^{\mu\nu}=\varepsilon u^{\mu}u^{\nu}
               -P\Delta^{\mu\nu} (u)
               +2W^{(\mu}u^{\nu )}
               +\pi^{\mu\nu} \label{eq:02},
\end{eqnarray}
where $\varepsilon\equiv T^{\mu\nu}u_{\mu}u_{\nu}$ is energy
density, $P\equiv -\frac{1}{3}\Delta_{\mu\nu} (u)T^{\mu\nu}$ is
pressure, $W^{\mu}\equiv
u_{\nu}T^{\nu\lambda}\Delta_{\lambda}^{\mu}(u)$ is energy flow and
$\pi^{\mu\nu}\equiv T^{\langle \mu \nu \rangle}$ is shear stress
tensor. The fluid is moving with four-velocity $u^{\mu}(x)$ such
that $u^{\mu} u_{\mu}=1$. We introduce transverse projection
tensor $\Delta^{\mu\nu}(u)\equiv g^{\mu\nu}-u^{\mu}u^{\nu}$ with
general metric $g^{\mu\nu}$ for the space-time considered. This
tensor has the property that $u_{\mu}\Delta^{\mu\nu}(u)
=u_{\nu}\Delta^{\mu\nu}(u)=0$. The notation used is:
$A^{(\mu}B^{\nu)}\equiv \frac{1}{2} (A^{\mu}B^{\nu} +
A^{\nu}B^{\mu})$ denotes symmetric tensor defined by two
four-vectors $A^{\mu}$ and $B^{\nu}$ and the angular bracket
notation $A^{\langle \mu \nu \rangle} \!\! \equiv \!\! \left[
\frac{1}{2}(\Delta^{\mu}_{\alpha}\Delta^{\nu}_{\beta} +
\Delta^{\mu}_{\beta} \Delta^{\nu}_{\alpha})
-\frac{1}{3}\Delta^{\mu\nu}\Delta_{\alpha\beta} \right]
A^{\alpha\beta}$ represents symmetric and trace free part of
tensor $A^{\mu\nu}$.

The net charge and entropy currents are generally given by,
respectively,
\begin{eqnarray}
 N^{\mu} &=& n u^{\mu},  \qquad    \label{eq:03} \\
 S^{\mu} &=& s u^{\mu} +\Phi^{\mu} \label{eq:04},
\end{eqnarray}
where $\Phi^{\mu}\equiv \Delta^{\mu}_{\nu}(u)S^{\nu}$ is entropy
flux. In this paper we shall use the so-called Eckart frame
\cite{Eckart1940} and fix arbitrary of choice of frame for the
four flow vector $u^{\mu}$. In what follows we assume that baryon
number, $N^{\mu}$, is conserved.\\

If the local thermal equilibrium is achieved (or, if the LRC is
switched off in our thought experiment and after the relaxation
process following it has been completed), the entropy flux
$\Phi^{\mu}$ should also disappear, its density reaches the
maximum, and it is locally conserved. The energy flow vector
$W^{\mu}$ and viscous shear tensor $\pi^{\mu\nu}$ should also
disappear. One gets therefore a perfect fluid, corresponding to
Eq.(\ref{eq:02}), (\ref{eq:03}) and (\ref{eq:04}), for which one
has
\begin{subequations}
\begin{eqnarray}
 T^{\mu\nu}_{\rm eq} &=& \varepsilon_{\rm eq}(T,\mu) U^{\mu}U^{\nu}
    -P_{\rm eq}(\varepsilon_{\rm eq},n_{\rm eq}) \Delta^{\mu\nu}(U),   \label{eq:05a} \quad \\
 N^{\mu}_{\rm eq} &=& n_{\rm eq}(T,\mu)  U^{\mu},                      \label{eq:05b} \\
 S^{\mu}_{\rm eq} &=& s_{\rm eq}(T,\mu)  U^{\mu},                      \label{eq:05c}
\end{eqnarray}
\end{subequations}
where $T=T(x)$ and $\mu=\mu(x)$ are the temperature and (baryonic)
chemical potential field given by $\varepsilon_{\rm eq}(x)$ and
$n_{\rm eq}(x)$. The equation of state $P_{\rm eq}(x)=P_{\rm
eq}(\varepsilon_{\rm eq}(x), n_{\rm eq}(x))$ for the perfect fluid
are assumed to be known. Here, $U^{\mu}(x)$ is the flow vector
appearing after relaxation has been completed.

\section{Relativistic hydrodynamics in the presence of Long-range correlations}\label{sec:03}

\subsection{Reference fluid field (RFF)}

Let us introduce reference fluid field (RFF), by which we
understand a set of solutions for a given perfect fluid without
the LRC (because LRC is assumed to be the origin of dissipation
and viscosity, one has perfect fluid if LRC are not present). In
our case the perfect fluid dynamics is given by specifying
\begin{subequations}\label{eq:06}
\begin{eqnarray}
 T^{\mu\nu}_{{\rm eq};\mu}&=&0, \label{eq:06a}\\
 N^{\mu}_{{\rm eq};\mu}&=&0,    \label{eq:06b}\\
 S^{\mu}_{{\rm eq};\mu}&=&0.    \label{eq:06c}
\end{eqnarray}
\end{subequations}
Note that there are $6$ unknown variables in the perfect fluid
hydrodynamics: energy density $\varepsilon_{\rm eq}(x)$, conserved
charge density, $n_{\rm eq}(x)$, pressure $P_{\rm eq}(x)$ and
three unknown components in the fluid velocity, $U^{\mu}(x)$. On
the other hand, there are $6$ equations: the first two equations
of Eq.(\ref{eq:06}) and equation of state, $P_{\rm eq}(x)=P_{\rm
eq}(\varepsilon_{\rm},n_{\rm eq})$ (equation Eq.(\ref{eq:06c}),
the local entropy conservations, can be derived from the previous two equations.).

\subsection{Effects caused by the LRC}

After switching on the LRC in our thought experiment (i.e.,
for a non-equilibrium dissipative situation) we have to solve
Eq.(\ref{eq:01}) with Eqs.(\ref{eq:02}), (\ref{eq:03}) and
(\ref{eq:04}). The pressure $P$ is different from the isotropic
equilibrium pressure, $P_{\rm eq}(\varepsilon_{\rm eq}, n_{\rm
eq})$, which can be obtained by equation of state (EoS) via the
equilibrium energy density, $\varepsilon_{\rm eq}$, and the baryon
number density, $n_{\rm eq}$ (i.e., $P \ne P_{\rm
eq}(\varepsilon_{\rm eq},n_{\rm eq})$).
The difference $P-P_{\rm eq}$ is usually regarded as a bulk pressure, $\Pi\equiv P-P_{\rm
eq}$. For energy density it is also natural to introduce bulk
energy density, $\Lambda\equiv \varepsilon-\varepsilon_{\rm eq}$,
because the microscopic occupation probability does not follow the
Boltzmann-Gibbs statistics any longer. Hence, the energy density
$\varepsilon$ in Eq.(\ref{eq:02}) cannot be expressed as a
function of the usual temperature $T$ and (baryonic) chemical
potential $\mu$ such like $\varepsilon_{\rm eq}(T,\mu)$.
Therefore, one can write thermodynamical quantities in the
presence of the long-range correlations,
\begin{subequations}\label{eq:07}
\begin{eqnarray}
 \varepsilon &=& \varepsilon_{\rm eq}(T,\mu) +\Lambda,    \label{eq:07a} \\
 n &=& n_{\rm eq} (T,\mu) +\delta n,                      \label{eq:07b} \\
 P &=& P_{\rm eq}(\varepsilon_{\rm eq},n_{\rm eq}) +\Pi.  \label{eq:07c}
\end{eqnarray}
\end{subequations}
\\

Now, the energy-momentum tensor Eq.(\ref{eq:02}) can be
decomposed by the flow vector $U^{\mu}(x)$ introduced
at the end of Sec. II, which is generally different from
$u^{\mu}(x)$. Denoting by $\delta u^{\mu}$ the difference between
$u^{\mu}(x)$ and $U^{\mu}(x)$, one can write
\begin{eqnarray}
 u^{\mu}(x) = U^{\mu}(x) + \delta u^{\mu}(x). \label{eq:08}
\end{eqnarray}
Since both $u^{\mu}$ and $U^{\mu}$ must satisfy normalization
conditions as physical flow vectors, i.e., $u^{\mu}u_{\mu}=1$ and
$U^{\mu}U_{\mu}=1$, they are related in the following way:
\begin{eqnarray}
\frac{1}{2}\delta u^{\mu} \delta u_{\mu} &=& -U^{\mu}\delta u_{\mu} \nonumber \\
 &=& u^{\mu}\delta u_{\mu} \equiv \gamma \label{eq:09}
\end{eqnarray}

It is desirable to describe the number of unknowns before moving
to the main task. In dissipative hydrodynamical model there are
$14$ unknowns: $\Lambda$, $\Pi$, $\delta n$, $3$ components of
$W^{\mu}$ (because condition $W^{\mu}u_{\mu}=0$), $5$ components
$\pi^{\mu\nu}$ (symmetric and traceless and satisfying conditions
$\pi^{\mu\nu}u_{\mu}=0$) and $3$ components of $\delta u^{\mu}$.
On the other hand, $\varepsilon_{\rm eq}$, $n_{\rm eq}$ and $3$
components of the $U^{\mu}$ (i.e., $5$ variables) and EoS, $P_{\rm
eq}=P_{\rm eq}(\varepsilon_{\rm eq},n_{\rm eq})$, are known for
the reference fields. As will be shown in Sec. III~D and IV~C,
the $3$ unknowns, $\Lambda$, $\Pi$, $\delta n$, and
the $8$ unknowns in $W^{\mu}$ and $\pi^{\mu\nu}$, total 11 unknowns
are function of the $3$ unknown of $\delta u^{\mu}$ and $6$ known
variables describing the RFF.
Note that the number of unknown three variables for $\delta
u^{\mu}$ coincides with the number of spatial dimensions. They
depend on the nature of the LRC (for example, if the correlations
are totally spatially isotropic for a rest frame of $U^{\mu}$, the
number of unknowns is reduced to unity.). Therefore, we need to
derive one equation from Eq.(\ref{eq:01a}) and Eq.(\ref{eq:01b})
(besides eqs.(\ref{eq:06a}), (\ref{eq:06b}) and EoS) to determine
one unknown. This is, as shown later, Eq.(\ref{eq:30a}) or
equivalently Eq.(\ref{eq:31}).

\subsection{Tensor decomposition by means of the RFF}

Introducing projection operator accompanying $U^{\mu}$,
 $\Delta^{\mu\nu} (U) (\equiv \hat \Delta^{\mu\nu})=g^{\mu\nu}-U^{\mu}U^{\nu}$,
one can decompose Eq.(\ref{eq:02}) as
\begin{eqnarray}
  T^{\mu\nu}=\hat \varepsilon~ U^{\mu} U^{\nu}
  -\hat P \Delta^{\mu\nu}(U)
  +2\hat W^{(\mu} U^{\nu )} +\hat \pi^{\mu\nu},
  \label{eq:10}
\end{eqnarray}
where $\hat\varepsilon$ and $\hat P$ are
\begin{subequations}
\begin{eqnarray}
\hat \varepsilon &\equiv&   T^{\alpha\beta} U_{\alpha} U_{\beta} = \varepsilon +3\Sigma, \label{eq:11a}\\
\hat P &\equiv&  -\frac{1}{3} T^{\alpha\beta} \Delta_{\alpha\beta}(U) = P + \Sigma,            \label{eq:11b}
\end{eqnarray}
\end{subequations} respectively.
The quantity $\Sigma$ that appears both in the Eq.(\ref{eq:11a}) and (\ref{eq:11b})
is explicitly given by
\begin{eqnarray}
\Sigma = \frac{1}{3}h
             (\gamma^2-2\gamma)
             -\frac{2}{3}\gamma'(1-\gamma)+\frac{1}{3}\gamma'' , \label{eq:12}
\end{eqnarray}
where $h \equiv \varepsilon+P$ is natural extension of the
equilibrium enthalpy $h_{\rm eq} \equiv \varepsilon_{\rm
eq}+P_{\rm eq}$ and
\begin{eqnarray}
\gamma'  \equiv W_{\mu} \delta u^{\mu},\quad
\gamma'' \equiv \pi_{\mu\nu}  \delta u^{\mu} \delta u^{\nu}.   \label{eq:13}
\end{eqnarray}

For the corresponding energy flow and shear viscosity tensor, we have
\begin{subequations}
\begin{eqnarray}
\hat W^{\mu} &\equiv&
     \hat\Delta^{\mu}_{\alpha} T^{\alpha\beta} U_{\beta} \nonumber \\
             &=& (1-\gamma)(hX^{\mu}+Y^{\mu})  - Z^{\mu}-\gamma'X^{\mu} -\gamma''U^{\mu}, \qquad
\label{eq:14a} \\
\hat \pi^{\mu\nu} &\equiv&
     \Big[ \frac{1}{2} (\hat\Delta^{\mu}_{\alpha} \hat\Delta^{\nu}_{\beta}
            + \hat\Delta^{\mu}_{\beta} \hat\Delta^{\nu}_{\alpha})
            -\frac{1}{3} \hat\Delta^{\mu\nu} \hat\Delta_{\alpha\beta} \Big]~
               T^{\alpha\beta} \nonumber \\
            &=& \pi^{\mu\nu}  +\gamma'' U^{\mu}U^{\nu}+\Sigma \hat\Delta^{\mu\nu}\nonumber \\
            && + ~h X^{\mu}X^{\nu}+2X^{(\mu}Y^{\nu)}  +2U^{(\mu}Z^{\nu)},
\label{eq:14b}
\end{eqnarray}
\end{subequations}
respectively, where
\begin{subequations}
\begin{eqnarray}
X^{\mu} &\equiv& \Delta^{\mu}_{\alpha}(U) u^{\alpha}  = \delta u^{\mu}+\gamma U^{\mu}, \label{eq:15a}\\
Y^{\mu} &\equiv& \Delta^{\mu}_{\alpha}(U) W^{\alpha}  = W^{\mu}+\gamma' U^{\mu},       \label{eq:15b} \\
Z^{\mu} &\equiv&  \pi^{\mu\nu}\delta u_{\nu}.                                          \label{eq:15c}
\end{eqnarray}
\end{subequations}
which are vectors that can be constructed from $\delta
u^{\mu}$, $W^{\mu}$ and $\pi^{\mu\nu}$. One can write then that
\begin{subequations}
\label{eq:hat_equations}
\begin{eqnarray}
 \hat \varepsilon =\varepsilon_{\rm eq} +\Lambda + 3\Sigma , \quad
 \hat P =P_{\rm eq} +\Pi + \Sigma.    \label{eq:16}
\end{eqnarray}
\end{subequations}

Here we approach the essence of our model.  In the case when
$\hat\varepsilon=\varepsilon_{\rm eq}(x)$ and $\hat P= P_{\rm
eq}(x)$ one has that
\begin{subequations}       \label{eq:17}
\begin{eqnarray}
\Lambda &=& -3\Sigma,      \label{eq:17a}\\
\Pi     &=&  -\Sigma.      \label{eq:17b}
\end{eqnarray}
Assuming now that \footnote{These limitations does not mean that
in general $W^{\mu}=0$ and $\pi^{\mu\nu}=0$.}
\begin{eqnarray}
\hat W^{\mu} &=& 0,        \label{eq:17c}\\
\hat \pi^{\mu\nu} &=&0,    \label{eq:17d}
\end{eqnarray}
\end{subequations}
one can link $T^{\mu\nu}$ to its equilibrium correspondence
$T^{\mu\nu}_{\rm eq}$. Physically, when the LRC is switched off
the flow velocity $u^{\mu}$ (and also the related thermodynamical
quantities) can not change into corresponding $U^{\mu}$ of the RFF
instantaneously. The flow vector field $u^{\mu}(x)$ approaches its
equilibrium limit $U^{\mu}(x)$ in a finite proper time. One may
regard such change as a kind of {\it relaxation process}. In the
next Section {\ref{sec:04}} we shall derive the corresponding
evolution of $\delta u^{\mu}$.\\

Conditions Eqs.(\ref{eq:17}) can be also mathematically expressed
by the following transformation:
\begin{eqnarray}
  {\cal M}[T^{\mu\nu}(u^{\mu})]  = T_{\rm eq}^{\mu\nu}(U^{\mu}).
 \label{eq:18}
\end{eqnarray}
Assumption (\ref{eq:18}) is natural because dissipative fluid must
correspond to some ideal fluid, which can be obtained by switching
off the LRC. It is our basic assumption deciding on the form of the
energy flow $W^{\mu}$ and shear viscosity $\pi^{\mu\nu}$, as will be
shown below.\\

\subsection{The model energy-momentum tensor}

The assumption discussed in the previous Section concerning
the energy-momentum tensor $T^{\mu\nu}$ determines also the
complete form not only of $\Pi$ but also both of $W^{\mu}$ and
$\pi^{\mu\nu}$. From Eqs.(\ref{eq:17a}) and (\ref{eq:17b}), we
obtain that  \footnote{The factor $3$ in the r.h.s. in
Eq.(\ref{eq:19}) comes from the number of spatial dimensions.}
\begin{eqnarray}
    \Lambda = 3\Pi , \label{eq:19}
\end{eqnarray}
i.e., that
\begin{subequations}
\begin{eqnarray}
\varepsilon &=& \varepsilon_{\rm eq}(T,\mu) +3\Pi, \label{eq:20a} \\
          P &=& P_{\rm eq}(T,\mu) +\Pi.            \label{eq:20b}
\end{eqnarray}
\end{subequations}
Proceeding further, conditions $\hat W^{\mu}\equiv 0$ and $\hat
\pi^{\mu\nu} \equiv 0$ determine the explicit form of the energy
flow vector and the shear tensor in the energy-momentum tensor in
Eq.(\ref{eq:01}). To give the $W^{\mu}$ and $\pi^{\mu\nu}$ correct
physical meaning of the, respectively, energy flow vector and shear
viscous tensor, we require that $W^{\mu}u_{\mu} =0$ (or
$W^{\mu}\delta u_{\mu}=\gamma'$) and that $\pi^{\mu\nu} u_{\nu}=0$
(or $\pi^{\mu\nu}\delta u_{\mu}=Z^{\nu}$).
The later condition can fix the form of the four vector $Z^{\mu}
=-\left[(1-\gamma)\Sigma+\gamma' \right]X^{\mu} -\gamma''
U^{\mu}$. The former condition $W^{\mu}u_{\mu}=0$ gives a relation
between $\Sigma$ and $\gamma'$. From the definition
$\gamma''\equiv Z^{\mu}\delta u_{\mu}$, eliminating $Z^{\mu}$ and
$\gamma'$, we can also find relation between $\gamma''$ and
$\Sigma$. When these expressions for $\gamma'$ and $\gamma''$ are
put into Eq.(\ref{eq:12}), then one finds that
\begin{eqnarray}
     \Sigma = \frac{\gamma(2-\gamma)}{(2\gamma-1)(2\gamma-3)} ~h . \label{eq:21}
\end{eqnarray}
Eliminating $\gamma',\gamma''$ and $Z^{\mu}$ from the equation
$\hat{W}^{\mu}=0$, one can obtain expression for $W^{\mu}$ in
terms of $U^{\mu}$ and $\delta u^{\mu}$. This result for $W^{\mu}$
put into the definition of $Y^{\mu}$ leads to $Y^{\mu}=-(h+\Sigma)
X^{\mu}$. Substitution $X^{\mu}$, $Y^{\mu}$, $Z^{\mu}$ and
$\gamma''$ into the equation $\hat{\pi}^{\mu\nu}=0$, one gets
expression for $\pi^{\mu\nu}$. Finally, one finds that the energy flow
vector and the shear viscous tensor have the following form:
\begin{subequations}
\begin{eqnarray}
W^{\mu} &=& -h_1\delta u^{\mu}
   -\left[ \gamma h_4-3\Sigma \right]U^{\mu},                     \label{eq:22a} \quad \\
\pi^{\mu\nu} &=&  \gamma^2 h_4  U^{\mu}U^{\nu}
  -\Sigma \Delta^{\mu\nu}(U)     \nonumber \\
  &+& h_2 \delta u^{\mu}\delta u^{\nu}
    -2\left[2\Sigma- \gamma h_4  \right]\delta u^{(\mu}U^{\nu )}, \label{eq:22b}
\end{eqnarray}
\end{subequations}
where $h_{n=1,2,4} \equiv h+n\Sigma$. Since $h \equiv
\varepsilon+P$, one obtains that $h=h_{\rm eq} +4\Pi$ (see
Eq.(\ref{eq:20a}) and Eq.(\ref{eq:20b})), where $h_{\rm eq}\equiv
\varepsilon_{\rm eq} +P_{\rm eq}$. Using this relation and
Eq.(\ref{eq:21}), one can write $\Pi (=-\Sigma)$ as function of
$\gamma$ and of equilibrium enthalpy $h_{\rm eq}$ as
\footnote{Note that we have following relation between $\gamma(x)$
and $\Lambda(x)$: $\gamma(x)= 1-\sqrt{1+\frac{\Lambda(x)}{h_{\rm
eq}(x)}}$. The sign of $\gamma(x)$ depends on the sign of
$\Lambda(x)$.}
\begin{eqnarray}
 \Pi = - \frac{1}{3}\gamma (2-\gamma) h_{\rm eq}(T,\mu). \label{eq:23}
\end{eqnarray}
By using Eq.(\ref{eq:23}), one can rewrite Eqs.(\ref{eq:22a}) and
(\ref{eq:22b}) in more compact form, such as
\begin{subequations}\label{eq:24}
\begin{eqnarray}
W^{\mu} &=& h_{\rm eq}(1-\gamma) \varphi^{\mu}, \label{eq:24a}\\
\pi^{\mu\nu} &=& h_{\rm eq} \varphi^{\mu} \varphi^{\nu} +\Pi \Delta^{\mu\nu}(u)
~=h_{\rm eq} \delta u^{\langle \mu} \delta u^{\nu \rangle}, \quad \label{eq:24b}
\end{eqnarray}
\end{subequations}
where
\begin{eqnarray}
 \varphi^{\mu} \equiv \gamma U^{\mu} -(1-\gamma) \delta u^{\mu}.
 \label{eq:25}
\end{eqnarray}
Hence, our model energy-momentum tensor is finally given by
\begin{eqnarray}
 T^{\mu\nu} 
 &=& [\varepsilon_{\rm eq}+3\Pi]u^{\mu}u^{\nu} -[P_{\rm eq}+\Pi]\Delta^{\mu\nu}(u) \nonumber \\
 &&+2h_{\rm eq} [1-\gamma]\varphi^{(\mu}u^{\nu)}
          +h_{\rm eq} \delta u^{\langle \mu} \delta u^{\nu\rangle}.  \qquad ~ \label{eq:26}
\end{eqnarray}
Since reference fields ($\varepsilon_{\rm eq},n_{\rm eq}, P_{\rm
eq}$ and $U^{\mu}$) are assumed to be known, the number of
unknowns in this model energy-momentum tensor, Eq.(\ref{eq:26}),
is only three (let us stress here that this is because of the Eq.(\ref{eq:18})).\\

It should be noted that expressions Eqs.(\ref{eq:24a}) and (\ref{eq:24b}),
one finds tensor relations
\begin{subequations}
\label{eq:27}
\begin{eqnarray}
 &&  W^{\mu}W_{\mu} = -3\Pi~(1-\gamma)^2 h_{\rm eq},  \label{eq:27a}\\
 && \pi^{\mu\nu}W_{\nu} = -2\Pi W^{\mu},              \label{eq:27b}\\
 && \pi^{\mu\nu}\pi_{\mu\nu} = 6\Pi^2,                \label{eq:27c}
\end{eqnarray}
\end{subequations}
which are same as found in the Non-extensive/dissipative correspondence \cite{OsadaPRC77} providing
\begin{eqnarray*}
 h_{\rm eq}[1-\gamma]^2 \leftrightarrow w_q [1+\gamma_q]^2,
\end{eqnarray*}
where $\gamma_q$ denotes the $\gamma$ that appear in
\cite{OsadaPRC77}.

\section{Evolution of $\delta u^{\mu}(x)$}\label{sec:04}

From the above discussion it is obvious that the most
important quantity to be discussed in detail is $\delta
u^{\mu}(x)$. We shall therefore derive here equation which $\delta
u^{\mu}$ should satisfy.

\subsection{Dissipation caused by the LRC}

Let us start with effects of dissipation caused by switching
on the LRC. Hereafter, we use , for simplicity, metric with
$g^{\mu\nu}_{;\mu}=0$. The relativistic hydrodynamical equation
Eq.(\ref{eq:01a}) can be always decomposed by contraction with
$u^{\mu}$ and $\Delta^{\lambda}_{\nu}(u)$ in the following way:
\begin{subequations}
\begin{eqnarray}\label{eq:28}
 &&u^{\mu} \partial_{\mu}\varepsilon +(\varepsilon+P)u^{\mu}_{;\mu}  \nonumber \\
 && \hspace*{15mm}  +W^{\mu}_{;\mu}+u_{\nu}[u^{\mu}W^{\nu}_{;\mu} +\pi^{\mu\nu}_{;\mu}
 ]=0,\qquad\label{eq:28a} \\
 && [(\varepsilon+P)u^{\mu}+W^{\mu}]u^{\lambda}_{;\mu}
 -\partial_{\mu}P\Delta^{\lambda\mu}(u)
 \nonumber \\
 && \quad
  + W^{\lambda}u^{\mu}_{;\mu}
  + [u^{\mu}W^{\nu}_{;\mu} +\pi^{\mu\nu}_{;\mu}
  ]\Delta^{\lambda}_{\nu}(u)=0.\qquad \label{eq:28b}
\end{eqnarray}
\end{subequations}
One can also write the corresponding hydrodynamical equations for
the RFF:
\begin{subequations}
\begin{eqnarray}
 &&U^{\mu} \partial_{\mu}\varepsilon_{\rm eq}
   +(\varepsilon_{\rm eq}+P_{\rm eq})U^{\mu}_{;\mu}
 =0,~\label{eq:29a}\\
 && (\varepsilon_{\rm eq}+P_{\rm eq})U^{\mu} U^{\lambda}_{;\mu}
 -\Delta^{\lambda\mu}(U)\partial_{\mu} P_{\rm eq}
 =0.~\label{eq:29b}
\end{eqnarray}
\end{subequations}
Inserting Eqs.(\ref{eq:24}) into Eqs.({\ref{eq:28a}}) and
({\ref{eq:28b}}) and then subtracting from them, respectively,
Eq.(\ref{eq:29a}) and (\ref{eq:29b}) one obtains
\begin{subequations}
\begin{eqnarray}
  && \gamma h_{{\rm eq};\mu}^{\mu}
   + \partial_{\mu} P_{\rm eq} \delta u^{\mu} +\Psi =0, \label{eq:30a} \\
  &&
   h_{{\rm eq};\mu}^{\mu} \varphi^{\lambda}
  + \partial_{\mu}P_{\rm eq}[ 2U^{(\mu}\delta u^{\lambda)}+\delta u^{\mu}\delta u^{\lambda}]
  + \Psi u^{\lambda} =0,  \qquad\quad \label{eq:30b}
\end{eqnarray}
\end{subequations}
where $h_{\rm eq}^{\mu}=h_{\rm eq}U^{\mu}$ is the enthalpy current
in the equilibrium state and $\Psi\equiv  h_{\rm eq}^{\mu}u_{\nu}
\delta u^{\nu}_{;\mu}$. By eliminating $\Psi$ from
Eq.(\ref{eq:30b}) using Eq.(\ref{eq:30a}) one finds that
Eq.(\ref{eq:30b}) is equivalent to Eq.(\ref{eq:29a}). This means
that Eq.(\ref{eq:30b}) is not independent equation. The unique
independent equation obtained from $T^{\mu\nu}_{;\mu} = 0$ with
model energy-momentum tensor (\ref{eq:26}) (besides
Eqs.(\ref{eq:29a}) and (\ref{eq:29b})) remains therefore
Eq.(\ref{eq:30a}), which can be expressed as a constitutive-like
equation:
\begin{eqnarray}
 \Pi+\tau_{\Pi}\dot{\Pi}&=&\frac{2}{3}(1-\gamma)
  \Big\{ \gamma \dot{\varepsilon}_{\rm eq}+\nabla_{\mu}P_{\rm eq}\delta u^{\mu}  \nonumber \\
 &&\qquad \qquad +h_{\rm eq}(\gamma \Theta + U^{\mu}\delta \dot{u}_{\mu})
  \Big\}~\tau_{\Pi},
 \label{eq:31}\quad
\end{eqnarray}
where $\dot{X}\equiv U^{\mu} X_{;\mu}$, $\nabla^{\mu}X \equiv
\Delta^{\mu\nu}(U)X_{;\nu}$ and $\Theta\equiv U^{\mu}_{;\mu}$. The
$\tau_{\Pi}$, which resembles a relaxation time, is given by
\begin{eqnarray}
\tau_{\Pi}\equiv -h_{\rm eq}(x)/\dot{h}_{\rm eq}(x) \label{eq:32}
\end{eqnarray}
and it is
determined by the property of the reference field $h_{\rm eq}(x)$,
i.e., by the equilibrium enthalpy. Because $\Pi$ is function of
$\delta u^{\mu}$, the Eq.(\ref{eq:31}) presents therefore the
condition that $\delta u^{\mu}$ should satisfy (as was also the
case for Eq.(\ref{eq:30a})).\\

Let us estimate $\tau_{\Pi}$ roughly, starting, for
simplicity, from 1+1 dimensional Bjorken scaling solution. The
enthalpy evolution is then given by the $h_{\rm eq}(\tau) \sim
(\tau_0/\tau)^{4/3}$, where $\tau$ is some proper time and
$\tau_0$ is initial proper time needed to equilibrate the fluid
without any long-range correlations. In this simple case, one
finds that Eq.(\ref{eq:32}) results in $\tau_{\Pi}=
\frac{3}{4}\tau$. It means that, if the fluid at the initial stage
is slightly different from the local equilibrium, it is difficult
to reach its thermalization as proper time passes. \\

In the more realistic 1+3 dimensional case, since the time
derivative of the enthalpy $|\dot{h}_{\rm eq}|$ in
Eq.(\ref{eq:32}) must be much lager than this simple estimation,
the $\tau_{\Pi}$ should be smaller than the previous estimation.
However, in this case the freeze out time, $\tau_{\rm fo}$, is
also shorter than in the 1+1 dimensional case. Therefore, whether
the stationary state (which is assumed to be create at the initial
stage in the relativistic heavy-ion collisions) can arrive to its
equilibrium depends on both the relaxation time and the freeze-out
time of the created matter. Note also that the relaxation time
$\tau_{\Pi}$ does not depend on $\delta u^{\mu}$. The
$\tau_{\Pi}$, in our model, is determined only by the nature of
the equilibrium state. On the other hand, the equilibrium time
$\tau_{\rm eq}$ should depend on the $\delta u^{\mu}$, which shows
how much away we are from the equilibrium, and it is obtained by
solving Eq.(\ref{eq:31}).

\subsection{A possible model for LRC}
We shall now discuss a possible model for the LRC in the
hydrodynamic approach. Notice that we have only three unknown
variables: $\delta u^{1}$, $\delta u^{2}$ and $\delta u^{3}$. This
is because $\Lambda$, $\Pi$, $W^{\mu}$ and $\pi^{\mu\nu}$ are
expressed in terms of $\delta u^{\mu}$ and RFF whereas $\delta
u^{0}$ can be determined by the Eq.(\ref{eq:09}). In Section IIIC
it is shown that, because of the charge conservation, $\delta n$
can be also written in terms of $\delta u^{\mu}$. Assuming that
the LRC are spatially isotropic $\delta u^{\mu}$ should also be
isotropic in the rest frame of RFF velocity $U^{\mu}$.\\

Let us consider a space-time point $x^{\mu}$ and the local
rest frame for the reference flow vector $U^{\mu}(x)=(1,0,0,0)$
there. Let us then suppose that in the rest frame the flow
velocity $u^{\mu}(x)$ is transformed in the following way:
\begin{eqnarray}
  && u^{\mu}(x) \mapsto v^{\mu}(x_0,\vec{x})=(\gamma_v,\gamma_v\vec{v}).
  \label{eq:33}
\end{eqnarray}
We use here the Minkowski metric instead of the general one
considered so far because it can be approximated with the flat
Minkowski metric $g^{\mu\nu}_{\rm m}=diag(1,-1,-1,-1)$ for
space-time sufficiently close to the space-time point $x$. If the
LRC introduced into perfect fluid is isotropic, the $\vec{v}$ is
also isotropic and $\vec{v}(\vec{x})=\vec{v}(|\vec{x}|)$, in the
rest frame of $U^{\mu}(x)$. Notice that $\vec{v}(\vec{x})$
fully characterizes the LRC used. In the case considered here the
local flow vector field $u^{\mu}(x)$ can be written as
\begin{eqnarray}
 u^{\mu}(x) = [\gamma_u, \vec{u}_{\parallel} +\vec{u}_{\bot}]
 \label{eq:34}
\end{eqnarray}
with $\gamma_u\equiv u^0=\sqrt{1-|\vec{u}_{\parallel} +\vec{u}_{\bot}|^2}$
and
\begin{eqnarray}
 \vec{u}_{\parallel}= \frac{\vec{v}_{\parallel}+\vec{U}}{1+ \vec{U}\cdot \vec{v}_{\parallel} }, \quad
 \vec{u}_{\bot} = \frac{\vec{v}_{\bot}}{\gamma_U[1+\vec{U}\cdot \vec{v}_{\parallel}]}.
 \label{eq:35}
\end{eqnarray}
Here, $\vec{v}_{\parallel}\equiv v_{\parallel} ~\vec{{\bf
e}}_{\parallel}$ and $\vec{v}_{\bot} \equiv v_{\bot} ~\vec{{\bf
e}}_{\bot}$ are, respectively, the parallel and perpendicular
components of $\vec{v}$ with respect to $\vec{U}$ whereas
$\vec{{\bf e}}_{\parallel}$, $\vec{{\bf e}}_{\bot}$ are their unit
vectors:
\begin{subequations}\label{eq:36}
\begin{eqnarray}
 v_{\parallel} 
 = \frac{\gamma_U(\vec{v}\cdot\vec{U})}{\sqrt{\gamma^2_U-1}} ~,\quad
 v_{\bot} = |\vec{v}| \sqrt{1-[\vec{\bf e}_{v}\cdot \vec{\bf e}_{\parallel} ]^2} ~,\quad
\end{eqnarray}
and
\begin{eqnarray}
 \vec{{\bf e}}_{\parallel} =  \frac{\gamma_U}{\sqrt{\gamma^2_U-1}}~\vec{U}, \quad
 \vec{\bf e}_{\bot} = \frac{ \vec{\bf e}_{v}-[\vec{\bf e}_{v}\cdot \vec{\bf e}_{\parallel}]
    \vec{\bf e}_{\parallel}}{\sqrt{1-[\vec{\bf e}_{v}\cdot \vec{\bf e}_{\parallel}]^2}}~.\qquad
\end{eqnarray}
\end{subequations}

Notice that the velocity field $u^{\mu}(x)$ (and, accordingly,
also $\delta u^{\mu}(x)$) are determined by the reference velocity
field $U^{\mu}(x)$ once we decide, by choosing $v(\vec{x})$, on
the type of the long-range correlations. As a possible example  of
the field $v(\vec{x})$, let us consider
\begin{eqnarray}
   v(\vec{x})= I(\vec{x}) \exp \left[ -\frac{ |\vec{x}|^2}{{L(x)}^2}\right],
   \label{eq:37}
\end{eqnarray}
where $I(x)$ and $L(x)$ are, respectively, the local correlation
intensity and correlation length. In what follows we shall assume
that only $I(x)$ varies with the expanding fluid whereas the $L$
is maintained as a global constant over the whole fluid evolution.
In other words, we assume the intensity $I(x)$ is determined by
the dissipative fluid dynamics whereas the correlation length $L$
is given by the particles composing the fluid. This assumption
fits nicely the ridge phenomenon found at RHIC discussed in
Section I, which indicates the persistency of the long-range
correlations during the fluid evolution.\\

To summarize this part, in our model its main quantity, $\delta
u^{\mu}$, is determined by only one equation Eq.(\ref{eq:30a}) (or
equivalently Eq.(\ref{eq:31})) and can be fixed by just one
variable, the correlation intensity $I(x)$. It means therefore
that correlation intensity itself is controlled by these equations
and that it could be identified with a kind of "relaxation" to the
equilibrium state taking place in the presence of LRC.

\subsection{Charge conservation}

Once $\delta u^{\mu}$ is fixed, $\Pi$ and $\Lambda$ are given by
Eq.(\ref{eq:19}) and (\ref{eq:23}), respectively. As for
$\delta n$, it is determined by the conservation of the (baryonic)
charge. For the conserved charge current $N^{\mu}$, we can write
explicitly that
\begin{subequations}
\begin{eqnarray}
 &&  \left[ (n_{\rm eq} +\delta n)(U^{\mu}+\delta u^{\mu}) \right]_{;\mu} =0, \label{eq:38a}\\
 &&  \left[n_{\rm eq} U^{\mu} \right]_{;\mu} =0.                              \label{eq:38b}
\end{eqnarray}
\end{subequations}
From the above two equations it is possible to write
\begin{eqnarray}
  \delta N^{\mu}_{;\mu}=0,
  \label{eq:39}
\end{eqnarray}
where $\delta N^{\mu} \equiv N^{\mu}-N^{\mu}_{\rm eq}$. Hence, the
correction of the (baryonic) charge density in the presence of LRC
is given by Eq.(\ref{eq:39}).\\

\section{Off-equilibrium entropy current}\label{sec:05}

For the equilibrium thermodynamical quantities one has
following fundamental thermodynamic relation,
\begin{eqnarray}
   T s_{\rm eq} = \varepsilon_{\rm eq} + P_{\rm eq} -\mu n_{\rm eq} , \label{eq:40}
\end{eqnarray}
which, using Eq.(\ref{eq:07}) and (\ref{eq:19}), one can be
rewritten as
\begin{eqnarray}
  s_{\rm eq}(T,\mu)=\frac{\varepsilon+P-\mu n}{T}-\left[ \frac{4\Pi-\mu\delta n}{T}\right].
\label{eq:41}
\end{eqnarray}
Notice that first term at the r.h.s. of Eq.(\ref{eq:41}) can be
regarded as a natural extension of the equilibrium entropy
density. We assume therefore that
\begin{subequations}
\begin{eqnarray}
  && s \equiv \frac{\varepsilon +P -\mu n}{T}, \label{eq:42a}
\end{eqnarray}
and, accordingly, that
\begin{eqnarray}
  && \delta s = \frac{4\Pi-\mu\delta n}{T}.    \label{eq:42b}
\end{eqnarray}
\end{subequations}

Although there is no principle to determine the explicit form of the  entropy flux
$\Phi^{\mu}$, a possible natural and simple candidate is
\begin{eqnarray}
  \Phi^{\mu}=\frac{W^{\mu}}{T} = \frac{h_{\rm eq}}{T}(1-\gamma) \varphi^{\mu}.
  \label{eq:43}
\end{eqnarray}
Using it the explicit form of the off equilibrium entropy
current is given by
\begin{eqnarray}
  S^{\mu} &=& (s_{\rm eq}+\delta s)u^{\mu}  +\Phi^{\mu} \nonumber \\
  &=& (1-\gamma) [ s_{\rm eq}U^{\mu}] 
  -\alpha \gamma [n_{\rm eq}U^{\mu}]-\alpha [\delta N^{\mu}]   +\Pi\beta^{\mu},  \quad \quad
  \label{eq:44}
\end{eqnarray}
where $\alpha\equiv \mu/T$ and $\beta^{\mu}\equiv \beta u^{\mu}$
with $\beta\equiv 1/T$. It should be noted here that the first
three terms of Eq.(\ref{eq:44}) are quantities of the first order
in $\gamma$, whereas the last term contains terms $\gamma^2$ (see
Eq.(\ref{eq:23}) and recall that $\gamma=-U_{\mu}\delta u^{\mu}$
is the first order infinitesimal quantity correction induced by
the LRC). It means that our model, in spite of applying simple
form of the entropy flux $\Phi^{\mu}$, introduces in a natural way
the $2^{nd}$ order terms, both in the energy-momentum tensor,
$T^{\mu\nu}$, and in the entropy current, $S^{\mu}$.  It is
interesting to notice that $\gamma$ can be related to a derivative
of the fluid velocity field $u^{\mu}$,
\begin{eqnarray}
 \delta u^{\mu} \sim u^{\mu}_{;\nu} \lambda^{\nu},
 \label{eq:45}
\end{eqnarray}
where $\lambda^{\mu}(x)$ is a kind of the local space-time scale.
It connects $\delta u^{\mu}$, which is crucial and most  important
parameter in our model, with the derivatives of the flow vector
$u^{\mu}$.\\

The second law of thermodynamics, $S^{\mu}_{;\mu}\geq0$, can be
expressed in our model as;
\begin{eqnarray}
 [\Pi \beta^{\mu}]_{;\mu}\geq
 (\partial_{\mu}\gamma)  h_{\rm eq}^{\mu} \beta
 +(\partial_{\mu}\alpha)[\gamma N^{\mu}_{\rm eq}+\delta N^{\mu}],
 \label{eq:46}
\end{eqnarray}
where current conservations, i.e., the second and third equations
of Eqs.(\ref{eq:06}) and Eq.(\ref{eq:39}) are used. In the baryon
free limit, one can also express the second law of thermodynamics
as an inequality of two kinds of time-like derivatives. Denoting
$\frac{dX}{d\tau}\equiv u^{\mu}X_{;\mu}$ and $\theta\equiv
u^{\mu}_{;\mu}$, one may write
\begin{eqnarray}
  \frac{d\Pi}{d\tau} + \frac{\Pi}{\tau_{\Pi}^*} \ge 
  -\frac{3}{2}\frac{1}{1-\gamma} \left[  \dot{\Pi} +\frac{\Pi}{\tau_{\Pi}} \right],
  \label{eq:47}
\end{eqnarray}
where $1/\tau_{\Pi}^*\equiv \theta-\frac{1}{T}\frac{dT}{d\tau}$.
The l.h.s. of Eq.(\ref{eq:47}) contains comoving time-like
derivative $d/d\tau$ with respect to the flow $u^{\mu}$ whereas
the r.h.s. contains comoving time-like derivative with respect to
the flow $U^{\mu}$ (denoted by dot). Note the minus sign in the
Eq.(\ref{eq:47}) which implies that in the case of weak LRC
(i.e., when $|\gamma|<1$), the inequality (second law of the
thermodynamics) holds providing that
\begin{eqnarray}
 \dot{\Pi} +\frac{\Pi}{\tau_{\Pi}} ~\ge 0 \quad\mbox{and}\quad
 \frac{d\Pi}{d\tau} + \frac{\Pi}{\tau_{\Pi}^*} ~\ge0.
\label{eq:48}
\end{eqnarray}
This requirement, which is the consequence of our model presented
here, restricts the relaxation process, namely the bulk pressure
$\Pi$ should decrease with proper time $\tau$ slower than
$\exp[-\tau/\tau^*_{\Pi}]$ in the rest frame of $u^{\mu}$ and
slower than $\exp[-\tau/\tau_{\Pi}]$ in the rest flame of
$U^{\mu}$.

\section{Summary and concluding remark}\label{Sec.Conclusion}

We have discussed dissipative effects appearing  by
introducing the long-range correlations (LRC) between particles
composing a perfect fluid. They arise because the perfectness of
the fluid is broken in such case, therefore such fluid obeys in
general equations of imperfect hydrodynamics. In particular, to
clarify the effects caused by the LRC, we have considered a
thought experiment in which LRC can be switch `on' and `off'. It
is assumed that the complete set of solutions for the perfect
fluid (called RFF) obtained when LRC are not present are known.
Then, the differences from the RFF, in velocity field, $\delta
u^{\mu}$, in pressure, bulk pressure $\Pi$, and in energy
density, $\Lambda$, can be regarded as resulting entirely from
dissipative effects caused by the LRC.\\

By switching off the LRC, the imperfect fluid should relax into a
perfect fluid. Such a relaxation process can be regarded as  a
transformation of the energy-momentum tensor. It is assumed to be
given by Eq.(\ref{eq:18}) and we have shown that all dissipative
terms in Eq.(\ref{eq:02}), such as $\Pi$, $W^{\mu}$ and
$\pi^{\mu\nu}$, are scaled by $\delta u^{\mu}$. Moreover, one of
dissipative equations Eq.(\ref{eq:01a}) can be expressed by
Eq.(\ref{eq:31}) (or, equivalently, by Eq.(\ref{eq:30a})).\\

It is noteworthy to observe that tensor relations
Eqs.(\ref{eq:27}) obtained in the presence of LRC, are the same as
those found in the dissipative fluid corresponding to the
non-extensive perfect fluid ($q$-fluid) discussed in
Ref.\cite{OsadaPRC77}.\\

When the LRC are spatially isotropic (in the rest frame of
$U^{\mu}$) with a constant correlation length $L$, we have shown
that the local correlation intensity $I(x)$ can be determined by
Eq.(\ref{eq:31}) which is equivalent to dissipative hydrodynamical
equation Eq.(\ref{eq:01a}). Interestingly, we found that it is
given by the form of constitutive-like equation. Moreover, it can
also be presented in our model as expression for the second law of
the thermodynamics, Eq.(\ref{eq:47}). The check of the causality
and stability of solutions obtained in our model as well as
numerical solutions  are beyond the scope of this brief paper and
will be discussed elsewhere.

We close with observation that our model can be applied to a
fluid with LRC originated from gravitational interaction.
The perfect hydrodynamics applied for the universe in the early stage
should be, for example, modified by such effects. However, this
problem should be considered using General Relativistic (GR)
hydrodynamics, as proposed in
\cite{Elze9809570,DemaretPRD21,SchutzPRD2}. The GR hydrodynamics
includes long-range interactions caused by gravity in a natural
way. If LRC originated by the gravity also cause dissipative
effects in a perfect fluid, then this may drive us to the
interesting question whether GR hydrodynamics have certain
correspondence in the (special) relativistic dissipative
hydrodynamics.\\

\begin{acknowledgments}
The author would like to thank Grzegorz Wilk for critical reading
of this manuscript.
\end{acknowledgments}

\end{document}